%% file: cas-dc-template.tex
\documentclass[a4paper,fleqn]{cas-dc}

\usepackage[numbers]{natbib}
\usepackage{CJK}
\usepackage{booktabs}
\usepackage{caption}
\usepackage{bicaption}
\usepackage{xcolor}
\usepackage{makecell}
\usepackage{url}
\usepackage{color,soul}
\usepackage{xspace}
\usepackage{graphicx,subfig}
\usepackage{diagbox}
\usepackage{rotating}

\newcommand{\dataset}{\textsc{MTips}\xspace}
\newcommand{\tool}{\textsc{GenTMS}\xspace}
\def\tsc#1{\csdef{#1}{\textsc{\lowercase{#1}}\xspace}}
\tsc{WGM}
\tsc{QE}
\tsc{EP}
\tsc{PMS}
\tsc{BEC}
\tsc{DE}

\begin{document}
\let\WriteBookmarks\relax
\def\floatpagepagefraction{1}
\def\textpagefraction{.001}

\shorttitle{Generating Tips from Song Reviews}
\shortauthors{Zang et~al.}

\title [mode = title]{Generating Tips from Song Reviews: A New Dataset and Framework}



%
\author[1]{Jingya Zang}[orcid=0000-0001-8720-7984]



\ead{20S051011@stu.hit.edu.cn}



\address[1]{organization={School of Computer Science and Technology, Harbin Institute of Technology}, city={Shenzhen},country={China}}

\author[1]{Cuiyun Gao}
\cormark[1]
\ead{gaocuiyun@hit.edu.cn}
\author[1]{Yupan Chen}
\ead{cyp36889@gmail.com}

\author[1]{Ruifeng Xu}
\ead{xuruifeng@hit.edu.cn}



\author%
[1]
{Lanjun Zhou}
\cormark[1]
\ead{bluejade.zhou@gmail.com}
\author%
[1]
{Xuan Wang}
\ead{wangxuan@cs.hitsz.edu.cn}

\cortext[cor1]{Corresponding author}



\begin{abstract}
Reviews of songs play an important role in online music service platforms. Prior research shows that users can make quicker and more informed decisions when presented with meaningful song reviews. However, reviews of music songs are generally long in length and most of them are non-informative for users. It is difficult for users to efficiently grasp meaningful messages for making decisions. 
To solve this problem, one practical strategy is to provide \textit{tips}, i.e., 
short, concise, empathetic, and self-contained
descriptions about songs. Tips are produced from song reviews and should express non-trivial insights about the songs.
To the best of our knowledge, no prior studies have explored the tip generation task in music domain. In this paper, we create a dataset named MTips for the task and propose a framework named \tool for automatically generating tips from song reviews.
The dataset involves 8,003 Chinese tips/non-tips from 128 songs which are distributed in five different song genres. 
Experimental results show that \tool achieves
top-10 precision at 85.56\%, outperforming the baseline models by at least 3.34\%. Besides, to simulate the practical usage of our proposed framework, we also experiment with previously-unseen songs, during which \tool also achieves the best performance with top-10 precision at 78.89\% on average. 
The results demonstrate the effectiveness of the proposed framework in tip generation of the music domain. 


\end{abstract}


\begin{highlights}
\item Propose a new task of tips generation from music reviews to facilitate users’ decision-making process. 
\item Summarize the characteristics of tips in music domain and release the first annotated Chinese dataset named \dataset for tip generation.
\item Present a novel framework named \tool which combines content-based ranking, stylistic ranking and diversity-weighted re-ranking to automatically produce tips from music reviews. 
\end{highlights}


\maketitle

\input{sections/Introduction}
\input{sections/dataset}
\input{sections/models}
\section{Conclusion and Future Work}
\label{sec:Conclusion}
In this paper, we are the first to propose a framework named \tool 
for generating tips from music reviews and publish the first Chinese tip dataset named \dataset in the music domain. 
We perform analysis of the dataset and make a synthesis of the effectiveness of each module of \tool.  
Benchmark results on top-k precision evaluation and practical evaluation show that the proposed framework 
is of high quality.
In the future, we will explore the use of tips for personalized recommendation. 
\bibliographystyle{cas-model2-names}

\bibliography{cas-refs}

\end{document}

%% file: sections/Introduction.tex
\section{Introduction}
Online music services, such as QQ Music, Pandora, Spotify, Apple Music, and Netease Music, etc., currently supply ever-growing catalogs with dozens of millions of music tracks. The services also provide entries for users to exchange their opinions about songs. The importance of song reviews has been proven for various aspects, such as music genre classification~\cite{DBLP:conf/ismir/OramasALSS16} and music recommendation~\cite{DBLP:conf/acl/TataE10}. 
\begin{figure}[h]
\centering
\includegraphics[width = 0.35\textwidth]{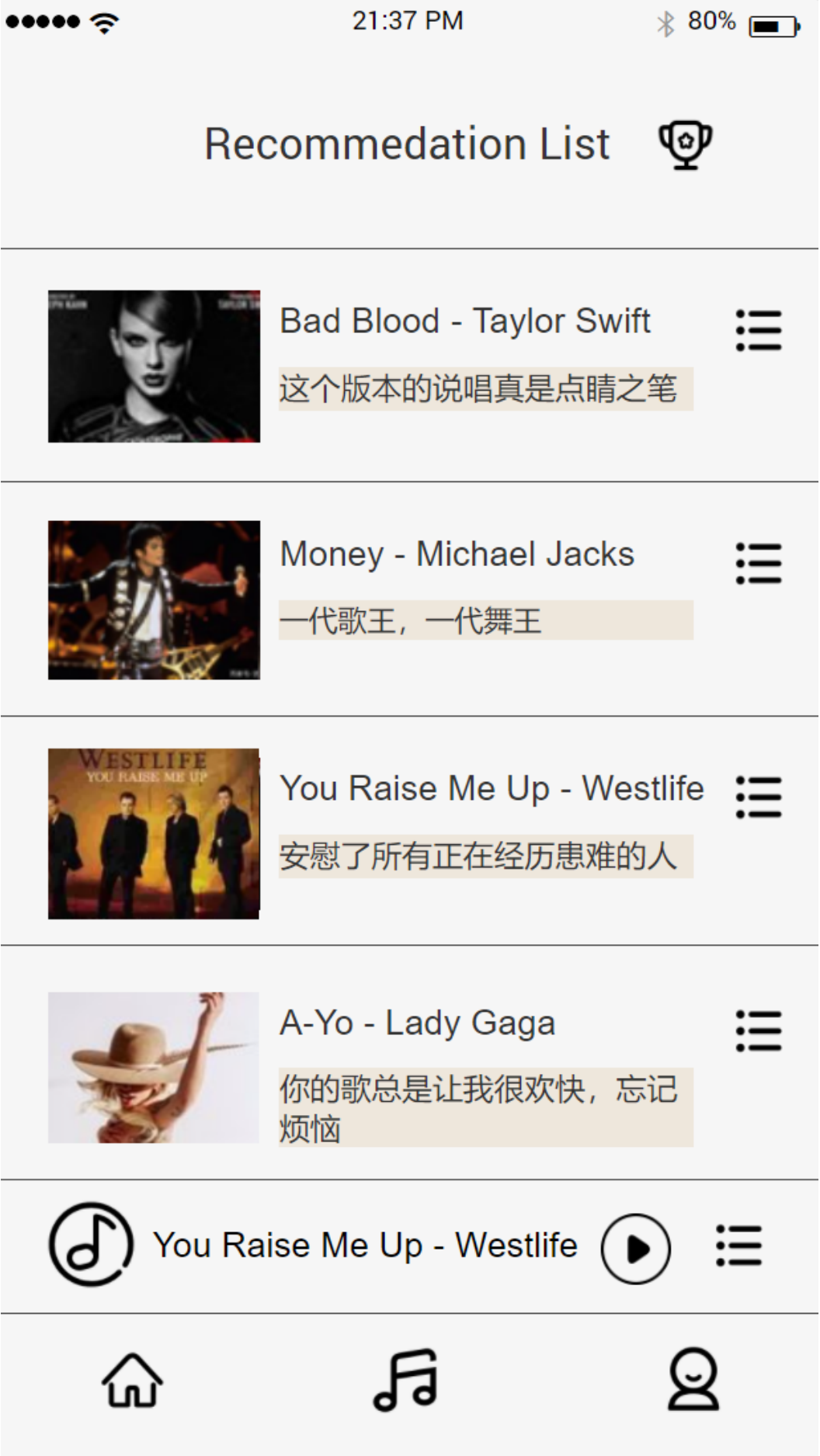}
\caption{Illustration of tips for music songs. Sentences in faint yellow background are tips. The tips are in Chinese. There is more than one tip for a song.}
\label{disp}
\end{figure}
Users can capture deeper content of a song, meet resonance, or discover interesting points from the reviews, beyond its title, attributes and description. Prior research~\cite{DBLP:conf/acl/TataE10} also shows that reviews facilitate users' decision-making processes in choosing which songs to listen to. 
However, reading a number of long-length reviews is a time-consuming process and the review
presentation 
space is limited for music platforms, especially on mobile devices~\cite{DBLP:conf/www/NovgorodovEGR19}.
One practical strategy is to provide \textit{tips} - concise, empathetic, and self-contained descriptions about the songs. Figure~\ref{disp} illustrates the tips for some music songs. Specifically, tips are review fragments or sentences, which provide insights about the melody, rhythm, context or affection of the songs. 
Considering the diverse perceptions of users, i.e., ``\textit{there are a thousand Hamlets for a thousand audiences}'', there generally exists several tips for one music song. 
To our best knowledge, no previous studies have explored the task of tip generation in music domain.

Although tip generation has been studied in other domains such as product~\cite{DBLP:conf/wsdm/HirschNGN21} and travel~\cite{DBLP:conf/www/GuyMNR17, DBLP:journals/dss/ZhuLZ18}, tips of music songs are different in the following two aspects: 1) they are more diverse in patterns since the characteristics, context, or affection of the song may be perceived by users variously; 
2) they are expected to be attractive and better with figures of speech such as metaphor, exaggeration, and rhyme, etc., based on the fact that no songs are essentially good or bad. Thus, the experience of tip generation in other domains is not applicable for music domain. 
Besides, the large number of reviews of music tracks, especially for popular songs, makes the process of finding useful information more challenging. Moreover, previous studies~\cite{DBLP:conf/naacl/GamzuGKLA21} generally extract short sentences from reviews by splitting based on full stops,
but the song reviews are often wrongly punctuated, e.g., the original review shown in Table~\ref{wrong punct}, resulting in long sliced sentences. 
Another splitting strategy is based on a slicing window, which may produce semantically-incomplete sentences,
e.g., ``\begin{CJK*}{UTF8}{gbsn}听着听着，就哭了，这首歌的MV我看过，\end{CJK*}''.

\begin{table*}[ht]
\renewcommand\arraystretch{1.2}
    \caption{An example for illustrating the punctuation correction result of our proposed model.
    The original punctuation and corrected one are highlighted
    in the second and last columns, respectively. The English translation is based on the corrected review. The bottom shows the sliced short sentences from original review and corrected review, respectively. The original review is sliced based on a slicing window; while the corrected review is split directly by full stops.
    }  
    \label{wrong punct}
\begin{CJK*}{UTF8}{gbsn}
\begin{tabular}{|c|ll|}
\hline
                                                                                                  & \multicolumn{1}{c|}{\textbf{Original Review}}                                                                                                                                                                                                                                                                     & \multicolumn{1}{c|}{\textbf{Corrected review}}                                                                                                                                                                                                                                               \\ \hline
\textbf{Chinese}                                                                                  & \multicolumn{1}{l|}{\begin{tabular}[c]{@{}l@{}}听着听着，就哭了\colorbox{yellow}{\textbf{\textcolor{red}{，}}}这首歌的MV我看过，男主角不\\ 帅，女主角不错\colorbox{yellow}{\textbf{\textcolor{red}{，}}}故事的最后令人感动。\end{tabular}} & \begin{tabular}[c]{@{}l@{}}听着听着，就哭了\colorbox{yellow}{\textbf{\textcolor{red}{。}}}这首歌的MV我看过，男主角不\\ 帅，女主角不错\colorbox{yellow}{\textbf{\textcolor{red}{。}}}故事的最后令人感动。\end{tabular} \\ \hline
\textbf{English}                                                                                  & \multicolumn{2}{l|}{\begin{tabular}[c]{@{}l@{}}After listening, I cried\textbf{\textcolor{red}{.}} I have seen the MV of this song. The leading actor is not handsome, but the leading\\ actress is beautiful\textbf{\textcolor{red}{.}} The end of the story is moving.\end{tabular}}                                                                                                                                                                                                                                                   \\ \hline
\multirow{5}{*}{\begin{tabular}{@{}c@{}}\textbf{Extracted} \\ \textbf{Short} \\ \textbf{Sentences}\end{tabular}} & \multicolumn{1}{c|}{\textbf{with Sliding Window}}                                                                                                                                                                                                                                                                 & \multicolumn{1}{c|}{\textbf{with Full Stops}}                                                                                                                                                                                                                                                \\ \cline{2-3} 
                                                                                                  & \multicolumn{1}{l|}{\begin{tabular}[c]{@{}l@{}}1.听着听着，就哭了，这首歌的MV我看过，\\ 2.就哭了，这首歌的MV我看过，男主角不帅，\\ 3.这首歌的MV我看过，男主角不帅，女主角不错，\\ 4.男主角不帅，女主角不错，故事的最后令人感动。\end{tabular}}                                                                                                                                                 & \begin{tabular}[c]{@{}l@{}}1.听着听着，就哭了。\\ 2.这首歌的MV我看过，男主角不帅，女主角不错。\\ 3.故事的最后令人感动。\end{tabular}                                                                                                                                                                                                \\ \hline
\end{tabular}
\end{CJK*}
\end{table*}

To address the above challenges, in this work, we first annotate a high-quality Chinese dataset containing tips of music songs, named \dataset.
To enhance the generalizability of the dataset, we involve reviews of songs from five different genres, including absolute music, rock and roll, the film \& television hit, Chinese pop, and Europe \& America \& Japan \& Korea (EAJK) pop. For ensuring the quality of the labeled data, we collaborated with one popular music platform\footnote{Due to the privacy issue, the company name is anonymized.}, and invited five annotators who are responsible for the corresponding business.
To correct the wrongly-used punctuation in user reviews, we propose a BERT-based punctuation prediction model. With the corrected punctuation, we can directly extract short sentences from reviews according to full stops. During manually labeling the short sentences, we summarize eight characteristics of music tips
categorized into two aspects, including the content relevance to the songs and stylistic pattern.
In total, we have labeled 8,003 Chinese tips/non-tips, involving 3,062 tips
from the top 100 reviews of 128 songs.



Based on the labeled dataset \dataset,
we then propose a novel framework, named \tool, for automatically
generating tips of music songs. \tool is built upon user reviews, and includes two major modules:

\textbf{(1)} Sentence relevance ranking module, which aims at scoring the representativeness of sentences for one song according to two aspects: the content relevance to the song and stylistic similarity to the annotated tips across songs. Besides the textual relevance, the content-based ranking module also involves the approval numbers of the reviews for the song, considering that the attribute reflects the degree of empathy delivered by the reviews. The stylistic-based ranking focuses on scoring whether the sentences share similar stylistic patterns as the annotated tips across songs.

\textbf{(2)} Diversity-weighted re-ranking module, which aims at increasing the diversity of the top-ranked tips for one song. Topic modeling and the combination of the two ranking scores are adapted in the module.

Experiments show that \tool can accurately generate the top-10 tips with precision 
score at 85.56\%. To simulate the practical usage of our framework, we also conduct experiment with previously-unseen 9 songs, achieving top-10 precision at 78.89\% on average.


The main contributions of this paper are as follows:
\begin{enumerate}
    \item To the best of our knowledge, we are the first to introduce and study the tip generation task in music domain. 
    \item We summarize the characteristics of tips in music domain and release the first annotated Chinese dataset named \dataset for tip generation for facilitating future research. We also provide detailed analysis of our dataset.
    \item We present a novel framework named \tool for automatically producing tips from music reviews. Extensive evaluation shows the effectiveness of our proposed framework.
\end{enumerate}

\textbf{Paper structure.} The remainder of the paper is organized as follows. Section~\ref{sec:Related Work} illustrates the related work. Section~\ref{sec:Experimental} introduces the annotation process and data analysis. Our proposed framework is presented in Section~\ref{sec:Methodology}. We describe the experiment setup in Section~\ref{sec:setup} and elaborate on the benchmark results in Section~\ref{sec:exper}. 
We conclude and mention future work in Section~\ref{sec:Conclusion}.

\section{Related work}
\label{sec:Related Work}
\subsection{Music Review Mining}
Mining music reviews has long been studied. Hu et al.~\cite{DBLP:conf/ismir/HuDWE05} demonstrate a system to predict star ratings by classifying reviews according to the genre of the music. In fact, in addition to star ratings, we can also assign text messages to music, which is more acceptable to users. 
Printer et al.~\cite{DBLP:conf/icwsm/PinterPSB20} create a dataset of expert reviews and offer several possible avenues for research. Different from their study in expert reviews, most of the work still focus on analyzing user reviews. Oramas et al.~\cite{DBLP:conf/ismir/OramasALSS16} explore a multimodal dataset constructed from albums and Amazon customer reviews. They analyze the potential correlation between sentiment and key culture events in music reviews.
Although prior researches have found music reviews useful for attracting potential users, 
no one studies the use of short sentences for the music songs, which is also known as ``Tips'' in the product domain~\cite{DBLP:conf/wsdm/HirschNGN21}. Previous studies show that ``Tips'' in the product domain is beneficial for item recommendation~\cite{DBLP:conf/cikm/YangHLWWZFHZW20}.

\subsection{Tip Generation in Product Domain}
Customer reviews contain valuable information which is useful for making a purchase decision. 
To help customers extract this information, researchers suggest new tasks of generating representative sentences from reviews for a given product. 
Gamzu et al.~\cite{DBLP:conf/naacl/GamzuGKLA21} propose an identification task on ``helpful sentence'' that is defined as individual sentence and can be helpful and faithful in a purchase decision.
Hirsch et al.~\cite{DBLP:conf/wsdm/HirschNGN21} define a tip in product domain as a short, concise, practical, and self-contained piece of advice, and publish an English dataset.
Several methods have been proposed for generating tips from user reviews.
For example, based on Transformer~\cite{vaswani2017attention}, Yang et al.~\cite{DBLP:conf/cikm/YangHLWWZFHZW20} propose a query-aware tip generation model for a popular e-commerce application, Dianping, but the dataset is not publicly available. 
Li et al.~\cite{DBLP:conf/www/LiWBL19, DBLP:conf/sigir/LiWRBL17} explore abstractive tip generation. 
However, they just enrich the representation of users and items with reviews in the training stage and do not introduce reviews to operating stage.
In this paper, we do not consider specific users but only focus on extracting representative tips to for a song.

\subsection{Text Summarization}
Text summarization is relevant to the tip generation,
which has been studied for a long time and is valuable to our task. 
Text summarization seeks to generate several sentences as a summary, which focuses on balancing the salience and redundancy of sentences. 
TextRank~\cite{DBLP:conf/emnlp/MihalceaT04} focuses on homogeneous graphs constructed with the content similarity among sentences. 
The heterogeneous graph shows its superiority by leveraging words as intermediate nodes~\cite{DBLP:conf/acl/WangLZQH20}.
Most of recent text summarization methods like BertSum~\cite{DBLP:conf/emnlp/LiuL19} encode the entire article, which is suitable for single-document-level summarization.
However, our task is to extract sentences from multiple reviews of a song, which is more similar to Multi-document Summarization (MDS) and the quantity of documents is not small.
Ma et al.~\cite{DBLP:journals/corr/abs-2011-04843} survey the deep learning techniques in MDS. The method of concatenation for multiple documents is still a research focus. 
Above methods focus on the salience of sentences and do not consider the redundancy. 
To mitigate the redundancy between extracted sentences, Zhou et al.~\cite{DBLP:conf/coling/ZhouWZ20} extract sub-sentential units based on the constituency parsing tree to construct a summary. 
Wu et al.~\cite{DBLP:conf/coling/YuanWL20} apply dependency analysis to sentences and propose an empirical method to split sentences into smaller units called ``fact''. However, these small units~\cite{DBLP:conf/acl/XuGCL20} decrease the readability of summary which is unacceptable in our task. 
Cho et al.~\cite{DBLP:conf/acl/ChoLFL19} present a novel method inspired by Determinantal Point Process  (DPP)~\cite{DBLP:journals/ftml/KuleszaT12} and capsule network to improve the similarity measure in extractive multi-document summarization. 
The redundancy, in other words, diverse content is also important when generating multiple tips for a song. 

%% file: sections/dataset.tex

\section{Data Annotation and Analysis}
\label{sec:Experimental}
In this section, we describe the workflow about the annotation process of Chinese tips for music songs. We obtained user reviews of songs from our business partner, one of the most popular Chinese music platforms. We first present the definition of the tips of songs, and then introduce the data preprocessing and annotation step. We finally summarize the characteristics of music tips, and conduct detailed analysis on the annotated dataset. 


\subsection{Definition about Tips of Songs}
The tips presented along with songs are aimed at attracting potential users, so they should strike a chord with users. Besides, due to the space limitation for display, as shown in Figure~\ref{disp}, tips should be short in length and convey clear information to users. Following Hirsch et al.~\cite{DBLP:conf/wsdm/HirschNGN21}, we define a \textit{tip} as a short, concise, empathetic, and self-contained sentence which is extracted from user reviews of a song. One song may possess several eligible tips. We restrict that the multiple tips identified for each song should be representative for diverse topics.

\subsection{Data Preparation}\label{subsec:data}
\begin{figure}[!t]
\centering
\includegraphics[width = 0.45\textwidth]{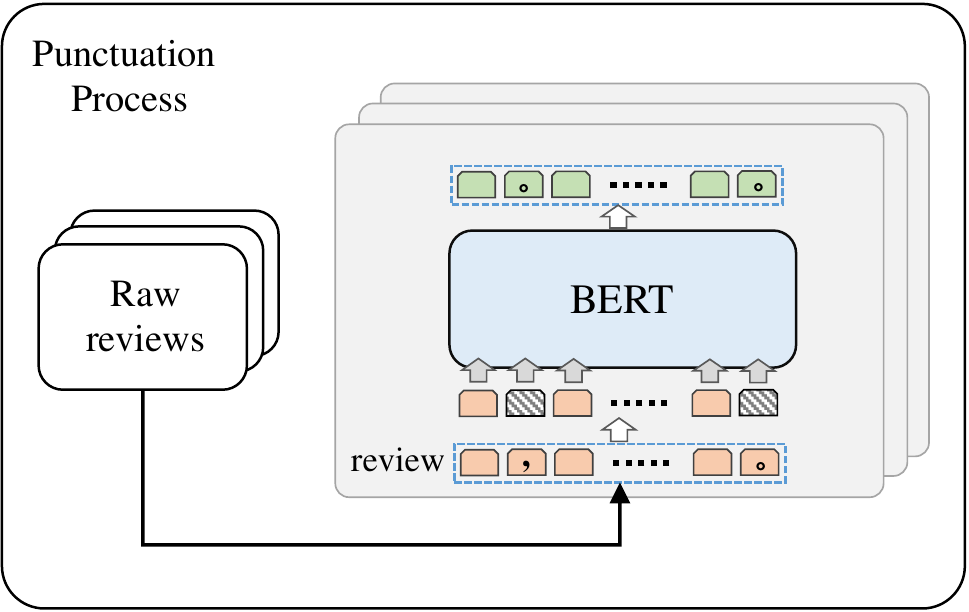}
\caption{The proposed BERT-based punctuation prediction model for review segmentation.}
\label{punctuate}
\end{figure}

In this section, we introduce how we prepare the data for subsequent manual annotation. 
We collect over 50,000 reviews from 128 songs which are distributed in five different song genres. 
During data annotation, we observe that the reviews are generally long in length (32.1 words per review on average in our collected corpus) and most of them are unmeaningful. To facilitate manual annotation, we propose several steps to select candidate sentences for the annotation. 
First, there are some duplicate words, misspellings, spaces, special symbols, etc., to be preprocessed. 
Second, we propose a novel process to split reviews into sentences. As can be seen in original Chinese reviews in Table~\ref{wrong punct}, the punctuation of user reviews is always confusing. Specifically, it is common for users to use a comma instead of a full stop at the end of a sentence.
To sovle that problem, we propose a BERT-based punctuation prediction model for more effective review segmentation. Figure~\ref{punctuate} shows the process.
We mask punctuation tokens in the input layer and predict them in the output layer. The number of classes in the output layer is set to 3, including full stop, comma, and empty string.
For training strategy, the vanilla BERT~\cite{DBLP:conf/naacl/DevlinCLT19} is pretrained with Wikipedia corpus, which has a large semantic gap with music review corpus. To mitigate this gap, we propose to fine tune the model on the prose corpus\footnote{https://github.com/Surprisezang/MTip.}. We use the prose corpus since the structure of prose is similar to reviews but the punctuation of prose is more rigid.
The classification accuracy of punctuation on the prose corpus is 93.7\%, indicating the effectiveness of punctuation prediction model. 
An example is shown in the column ``with Full Stops'' in Table~\ref{wrong punct}. Based on the BERT-based punctuation prediction model, we segment reviews into sentences for the subsequent annotation.

For facilitating manual annotation, we also provide the attributes
of songs, such as song names, singer names, lyrics, the number of approval, play count, movie description and song description. The details can facilitate the manual annotation process.

\begin{table*}[!t]
\caption{Criteria summary of annotated tips.}  
\label{table: Types of positive cases}
\centering
\scalebox{1.0}{
\begin{tabular}{c|c|c}
\toprule[1.5pt]
& Characteristic & Example \\
\midrule[1pt]
\multirow{8}{*}{Content aspect}  & \makecell[c]{Related to singer and lyricist} & \makecell[c]{\begin{CJK*}{UTF8}{gbsn}依然喜欢梁静茹清澈透明的声音  \end{CJK*} \\ (I still like Liang Jingru's clear and transparent voice)}\\
\cline{2-3}
& \makecell[c]{Related to lyrics and song title}& \makecell[c]{\begin{CJK*}{UTF8}{gbsn}我想去成都，去看看这个温柔的城市\end{CJK*} \\ (I really want to go to Chengdu to see the gentle city)}\\ 
\cline{2-3}
& \makecell[c]{Related to songwriting}
& \makecell[c]{\begin{CJK*}{UTF8}{gbsn}歌词令人警醒，无愧于十一项提名\end{CJK*} \\ (The lyrics are sobering, worthy of eleven nominations)}\\
\cline{2-3}
 & \makecell[c]{Background knowledge related to songs} & \makecell[c]{\begin{CJK*}{UTF8}{gbsn}哆啦A梦是我的童年回忆\end{CJK*} \\ (Doraemon is really a memory of my childhood)}\\
\midrule[0.5pt]
\multirow{8}{*}{Stylistic aspect} & Sensory artistic conception & \makecell[c]{\begin{CJK*}{UTF8}{gbsn}好甜，有恋爱的感觉\end{CJK*} \\ (It is so sweet and makes me feel like in love)}\\
\cline{2-3}
 & Scene artistic conception & \makecell[c]{\begin{CJK*}{UTF8}{gbsn}我说他日定会笑着相逢\end{CJK*} \\ (I said I would meet you with a smile one day)}\\
\cline{2-3}
 & \makecell[c]{Humorous description} & \makecell[c]{\begin{CJK*}{UTF8}{gbsn}每次听这首歌我都把油门踩到底了\end{CJK*} \\ (Every time I listen to this song I put my foot on the gas)}\\
\cline{2-3}
 & Philosophy and inspiration & \makecell[c]{\begin{CJK*}{UTF8}{gbsn}勇敢面对生活会让人生更有意义 \end{CJK*} \\ (The courage to face life can make your life more meaningful)}\\
\bottomrule[1.5pt]
\end{tabular}}
\end{table*}

\begin{figure}[ht]
\centering
\includegraphics[width = 0.45\textwidth]{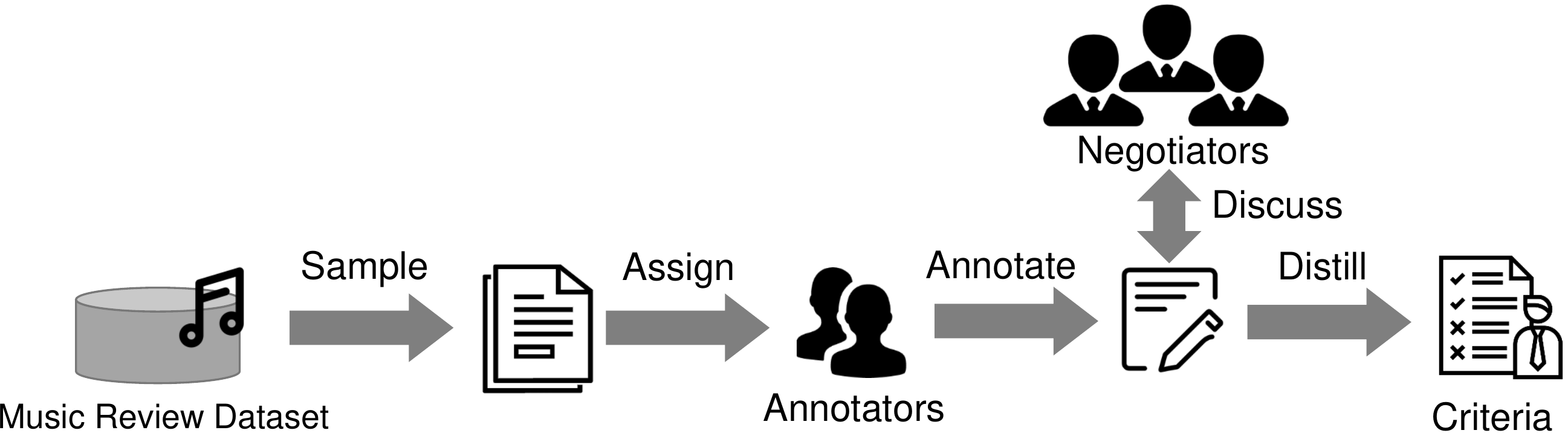}
\caption{Overview of the criteria annotation process.
}
\label{annotate}
\end{figure}

\subsection{Tip Annotation and Specification}\label{specification}
To ensure the quality of the annotated dataset, five company staff who are responsible for the tip generation project are invited to annotate the data. They are generally more familiar with user behavior and more careful during annotation than common users, therefore the quality of the labeled dataset can be guaranteed.
Before annotating all the sentences, we first need to determine the detailed criteria of tips for guiding the annotation. Figure~\ref{annotate} illustrates the criteria discussion process. We sample 2,000 candidates from the remained short sentences in Section~\ref{subsec:data}, and assign them to two annotators. The two annotators separately mark the candidates as tips or not, and briefly clarify the reason for the marking. After the labeling, the other three annotators join as negotiators to discuss the samples with divergent opinions. The sampled candidates are finally received a consensus among the annotators. The specification involves two aspects, i.e., content aspect and stylistic aspect. The detailed characteristics
of tips under the two aspects are summarized as follows, with examples illustrated in Table \ref{table: Types of positive cases}.


\subsubsection{Content Aspect}
The content of tips is generally related to the attributes of annotated songs. We summarize the content-related features of the tips into four characteristics.

(1) Related to singers and lyricists: This type of tips discusses about the singers or lyricists associated with songs, such as the singer's singing skill and lyricist's creative style. Table~\ref{table: Types of positive cases} gives an example, i.e., ``\textit{I still like Liang Jingru\'s clear and transparent voice}'' which is related to the singer's voice style.

(2) Related to lyrics and song titles: The type of tips is relevant to the lyrics or song titles of songs, since some lyric or song title can be classic for one song.  
For example, the tip ``\textit{I really want to go to Chengdu to see the gentle city}'' for the song ``Chengdu'' is associated to the song title.

(3) Related to songwriting of the songs: 
This type of tips related to the composition and status of a song. This is macro evaluation. 
For example, ``\textit{The lyrics are sobering, worthy of eleven nominations}'' speaks highly of the song ``Alright''.

(4)
Background knowledge related to songs: 
This involves other background knowledge related to songs, including associated movie, animations, etc. An example of ``Doraemon is really a memory of my childhood'' brings feelings about animations into songs.

\subsubsection{Stylistic Aspect}
Besides the content aspect, the annotated tips generally present similar stylistic patterns. We summarize the stylistic patterns into four characteristics.

(1) Sensory artistic conception: These tips mainly involve descriptions about abstraction and migration of feelings for a song. The descriptions can arouse sympathy among listeners. For example, ``\textit{It is so sweet and makes me feel like in love}'' expresses the sweet feeling in the heart.

(2) Scene artistic conception: Unlike (1), this kind of tips are related to a concrete scenario imagined by the listener, describing about the characters and events.
For example, ``\textit{I said I would meet you with a smile one day}'' is a review sentence for the song ``Lover's prattle'', depicting the day of reunion. Such tips can trigger associations from listeners, thus rendering the song attractive.

(3) Humorous description: The type of tips is characterized by humorous expression, which may be interesting for other listeners. Such tips generally contain unexpected or witty words, implying the content or rhythm of a song. For example, the tip ``\textit{Every time I listen to this song I put my foot on the gas}'' indicates the fast rhythm of the song.

(4) Philosophy and inspiration: The kind of tips expresses the implication of one song or philosophy of life. Such tips can be enlightening for listeners.
For instance, ``\textit{The courage to face life can make your life more meaningful}'' for the song ``Sailor'' is related to the song content and also inspiring for listeners.

\begin{figure}[ht]
\centering
\includegraphics[width = 0.45\textwidth]{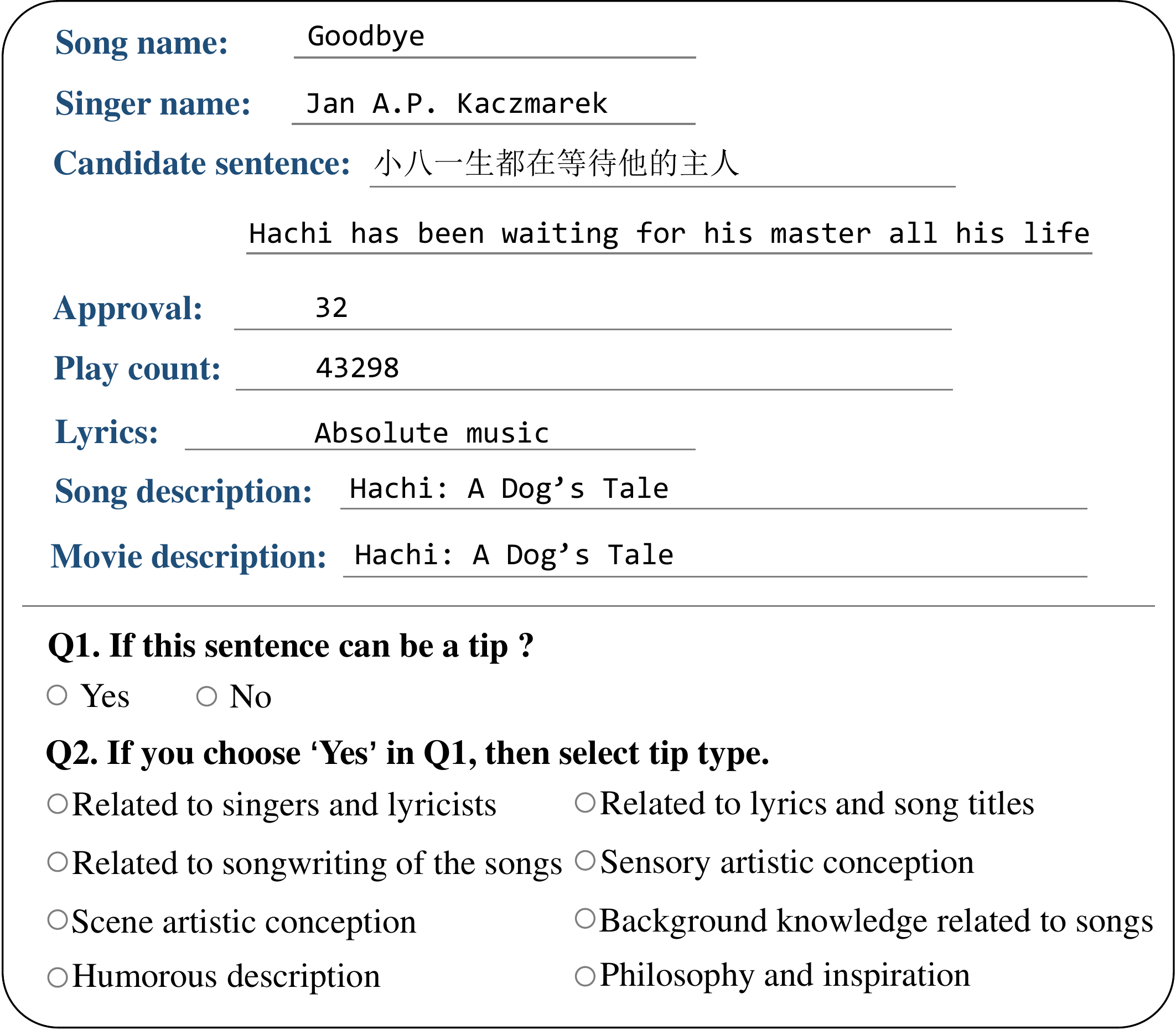}
\caption{An example of candidate for annotation.}
\label{example}
\end{figure}

The tips should satisfy at least one criterion for each of the two aspects, as shown in Table~\ref{table: Types of positive cases}.
Specifically, four annotators participate in the data labeling while the remaining one, serving as an inspector, is responsible for supervising the whole process. Figure~\ref{example} shows one example of candidate for annotation. Each of the four annotators is provided with the relevant information of one candidate and asked to choose whether the candidate is tip or not. To increase the credibility of the annotation, they also need to mark which characteristics the tips have. The Fleiss' Kappa score among the four annotators is 0.78, indicating a substantial agreement. Based on discussion with the inspector, all the candidates are finally labeled without disagreement. Overall, the 8,003 labeled samples contain 3,062 tips and 4,941 non-tips.



\begin{table*}[!t]
\normalsize
\caption{Distribution of tip characteristics across different genres of songs.}  
\label{dataset statistics}
\centering
\scalebox{1.0}{
\begin{tabular}{c|ccccc|c}
\toprule[1.5pt]
\textbf{Type of tips} & \makecell[c]{\textbf{Absolute}\\ \textbf{music}} & \makecell[c]{\textbf{Rock }\\ \textbf{and roll}} & \makecell[c]{\textbf{Film and}\\ \textbf{ television hit}} & \makecell[c]{\textbf{Chinese }\\ \textbf{pop}} & \makecell[c]{\textbf{EAJK}\\ \textbf{pop}} &  \makecell[c]{\textbf{Percentage}\\ \textbf{(\%)}} \\
\midrule[1pt]
Related to songwriting  &  2.19 & 1.14 & 0.49 & 2.22 & 3.23 & 9.27 \\
Related to singers and lyricists  & 0.42 & 1.70 & 1.37 & 8.13 & 4.57 & 16.19 \\
Related to lyrics and song titles  & 1.04 & 1.89 & 0.39 & 17.56 & 1.47 & 22.35 \\
Background knowledge related to songs &  4.24 & 2.68 & 5.75 & 2.48 & 3.79 & 18.94 \\
Sensory artistic conception & 1.47 & 1.18 & 0.65 & 4.54 & 2.45 & 10.29  \\
Scene artistic conception & 1.47 & 1.93 & 2.02 & 4.83 & 2.09 & 12.34 \\
Humorous description & 1.34 & 0.72 & 0.52 & 0.59 & 2.58 & 5.75 \\
Philosophy and inspiration & 0.78 & 0.65 & 0.20 & 2.81 & 0.43 & 4.87 \\
\hline
Song genre distribution & 12.95 & 11.89 & 11.39 & 43.16 & 20.61 &100.00\\
\bottomrule[1.5pt]
\end{tabular}
}
\end{table*}

\begin{table*}[htb]
\normalsize
\caption{Proportion of each Genre in Tips. EAJK is short for Europe, America, Japan and Korea.}  
\label{analysis}
\centering
\scalebox{1.0}{
\begin{tabular}{cccccc}
\toprule[1.5pt]
 & Absolute music & Rock and roll & The film and television hit & Chinese pop & EAJK pop\\
\midrule[1pt]
Avg. review len &28.34&28.52&35.60&42.63&32.30\\
Avg. tip len &14.00&13.02&13.72&13.42&12.88\\
\bottomrule[1.5pt]
\end{tabular}}
\end{table*}

\subsection{Analysis of Annotated Dataset}
\label{sec:analysis_dataset}
Table~\ref{dataset statistics} describes the characteristic distribution of the tips across
different genres of songs in the annotated dataset. As can be observed, the characteristic distribution for different genres of songs are different.
The most common types of tips are ``\textit{related to lyrics and song titles}'', ``\textit{background knowledge related to songs}'', and ``\textit{related to singers and lyricists}'', which is reasonable. For example, lyrics and song titles generally reflect the content of songs, so the related review sentences can deliver condensed information about the songs, which is helpful for attracting potential users. For the tips about ``\textit{background knowledge related to songs}'', we can find that they account for large proportions for the songs belonging to film \& television hit (5.75\%) and absolute music (4.24\%). This indicates that the background knowledge of songs such as films and animations is useful for listeners. Regarding to the tips about ``\textit{related to singers and lyricists}'', they occupy huge percentages in songs belonging to Chinese pop and EAJK pop, 8.13\% and 4.57\%, respectively. We suppose that listeners of the songs tend to post reviews about their favorite pop stars. Besides, the tips describing about ``\textit{humorous description}'' and ``\textit{philosophy and inspiration}'' present the lowest percentages among all the types, i.e., 5.75\% and 4.87\%, respectively.
We analyze the average numbers of words for the original reviews and tips, respectively. As shown in Table~\ref{analysis}, we find that the original reviews are significantly longer than tips with respect to the number of words. Generally, tips are described using around 13 words.



%% file: sections/models.tex
\section{Models}
\label{sec:Methodology}
\begin{figure*}[ht]
\centering
\includegraphics[width = 1.0\textwidth]{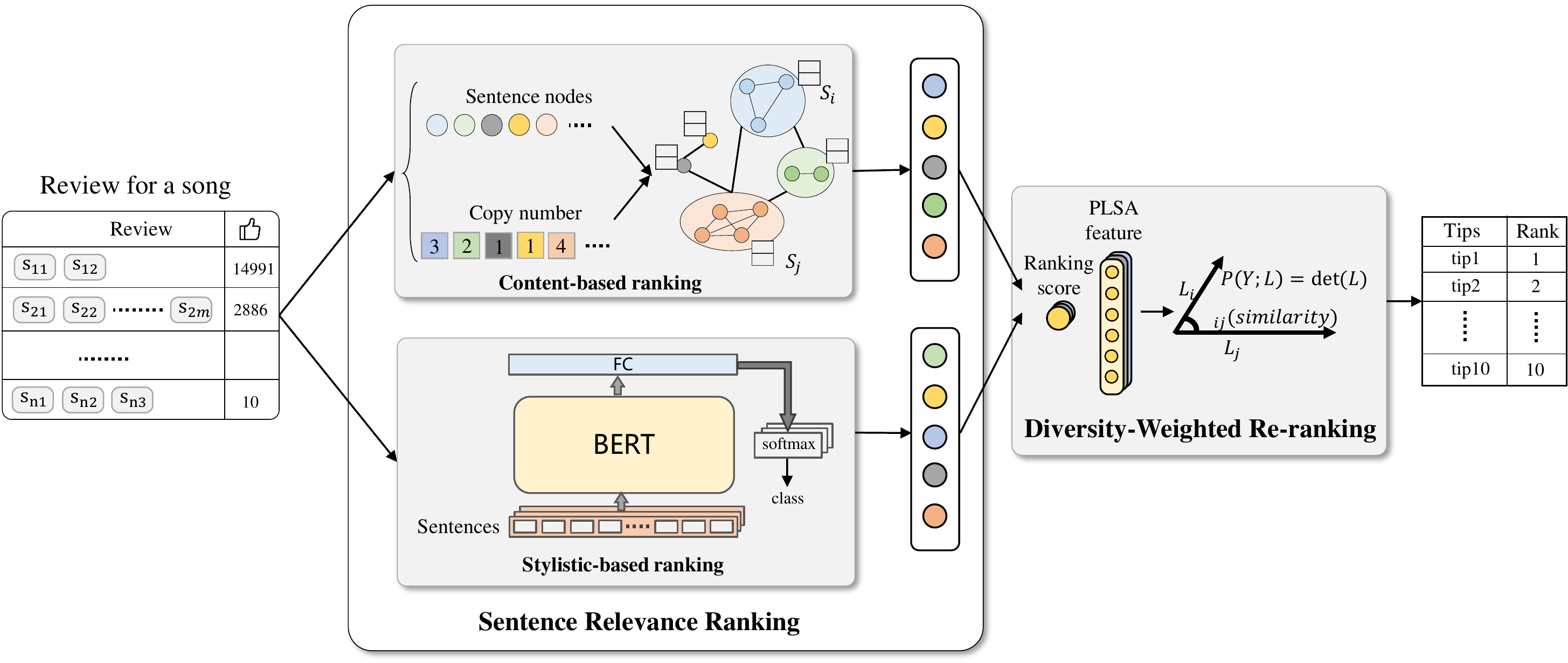}
\caption{Overview of the tip generation \tool. }
\label{data process}
\end{figure*}

In this section, we elaborate on the proposed framework \tool for automatic tip generation. Figure~\ref{data process} illustrates the workflow of \tool, including two main components: 
sentence relevance ranking, and diversity-weighted re-ranking. Given the candidate review sentences
for a song, sentence relevance ranking module is divided into two parts, content-based ranking and stylistic-based ranking. 
The content-based ranking component aims at measuring the content representativeness of the sentences within the reviews of a song. 
The stylistic-based ranking component focuses on shared stylistic patterns among tips across songs.
The diversity-weighted re-ranking component further adjusts the ranking scores by considering the topical diversity of selected tips. The output top-ranked sentences are regarded as tips.


\subsection{Sentence Relevance Ranking}
In this section, we propose two ranking methods which have different textual clues and play two roles in identifying tips. One is Content-based Ranking, which assumes that there are representative contents in the reviews of each song respectively. Another is Stylistic-based Ranking, which aims at identifying shared patterns across different songs.
\subsubsection{Content-based Ranking}
\begin{figure}[ht]
\centering
\includegraphics[width = 0.4\textwidth]{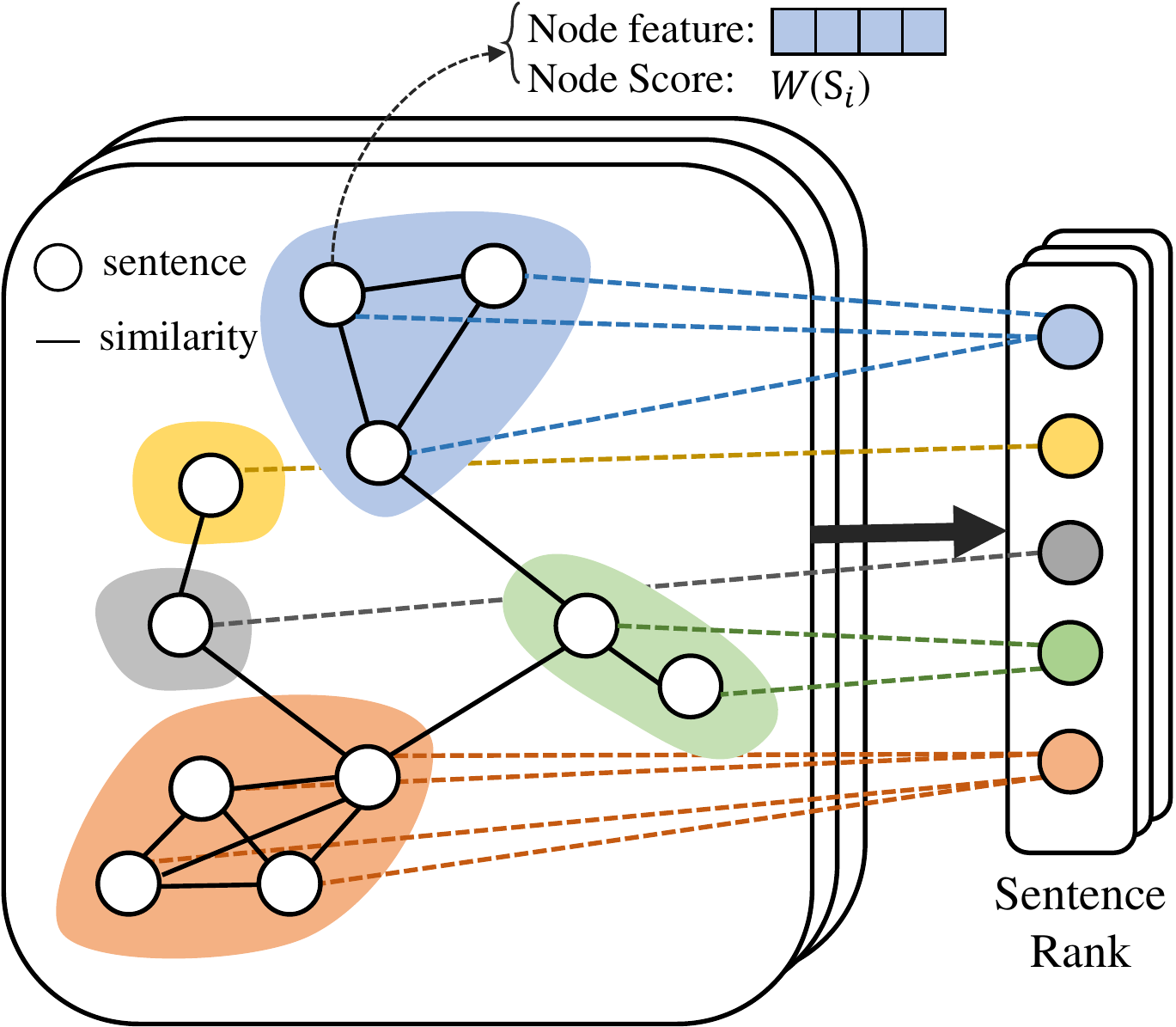}
\caption{Content-based ranking. $k = \log_m b$. $m$ is the base number and $b$ is approval number.}
\label{ltr}
\end{figure}

To prioritize representative sentences as tips, we propose a novel text ranking component based on TextRank~\cite{DBLP:conf/emnlp/MihalceaT04}. In this component, we take the reviews of a song as a document and rate the sentences in an unsupervised way.
We construct a sentence graph, as shown in Figure~\ref{ltr}, in which each node represents a sentence and the edges are weighted by
the word bag similarities between associated nodes. 
Let $\tilde{D}=\{{\tilde{S}_1}, {\tilde{S}_2}, ..., {\tilde{S}_N}\}$ denotes the review of a song which contains $N$ sentences, where $\tilde{S}_i$ is the $i$-th sentence. 
$B=\{{b_1}, {b_2}, ..., {b_N}\}$ is the corresponding approval numbers of each sentence. Considering that reviews with higher approval numbers are more likely to be representative and attractive for users,
we incorporate this property into TextRank. 
To involve the approval number $b$, we propose to repeat the sentence nodes $k$ times where $k=\log_{m} {b}$ and $m$ denotes the hyper-parameter for scaling the approval numbers.
The node repetition is proposed to rank the sentences considering
the approval numbers.
The new sentence set is $D=\{S_1, S_2, ..., S_M\}$, and $M \geq N$. 
Let $S_i=\{w_{1}, w_{2}, ..., w_{T}\}$ denote $T$ words in sentence $S_i$.
There is an edge between two sentence nodes if the two sentences have a common word.
The edge weight is defined as:
\begin{equation}\label{eq6}
e(S_i, S_j) = \frac{ \left\lvert{w_k \vert w_k \in S_i \& w_k \in S_j} \right\rvert}{log(|S_i|)+log(|S_j|)},
\end{equation}
where $e(S_i, S_j)$ denotes the weight between $S_i$ and $S_j$, and $w_k$ is the $k$-th word in the sentence. $log(|S_i|)$ is the length penalty on $S_i$, and $|\cdot|$ denotes the word length of $\cdot$.

The node weight is defined as:
\begin{small}
\begin{equation}\label{transfer eq}
\begin{aligned}
W(S_{i}) = \frac{(1-d)}{M} + d*\sum_{S_j \in E(S_i)}{\frac{e(S_j, S_i)}{\sum_{S_k \in E(S_j)} {e(S_j, S_k)}}W(S_{j})},
\end{aligned}
\end{equation}
\end{small}

\noindent where $E(S_i)$ denotes the sentence set with edges in $S_i$.
The second term in the equation means a sentence may affect the other sentences with edge connection, where $d$ is a damping factor controlling the impact from the relation with other sentences. 
Notice that there are duplicated sentences in the graph. 
The content-based ranking score $Score_a(\tilde{S})$ for the unique sentence $\tilde{S}$ is defined as below:

\begin{equation}\label{eq7}
\begin{aligned}
Score_a(\tilde{S}) = \sum_{\tilde{S}=S_j}{W(S_j)}, \qquad j=1,2,...M.
\end{aligned}
\end{equation}




\subsubsection{Stylistic-Based Ranking}
As aforementioned in Section~\ref{specification}, tips often present
stylistic patterns to attract users. 
This type of pattern is song-independent and can be captured across songs. 
In this component, we use our dataset to train a song-unaware classifier in a supervised way. 
We adopt the pretrained language model to modeling common patterns 
of tips across songs. 
First, we obtain the representations of sentences.
Formally, for a candidate sentence
$\tilde{S_i}=\{w_{1}, w_{2}, ..., w_{T}\} \in D$, we feed it into BERT and obtain
hidden states:
\begin{small}
\begin{equation}\label{inter1}
\begin{aligned}
h_{[CLS]}, h_1, ..., h_T, h_{[SEP]}=BERT([CLS], w_1, ..., w_T, [SEP])
\end{aligned}
\end{equation}
\end{small}
\noindent where the special [CLS] token is added to the front of a sequence and denotes the start of a sentence, and [SEP] denotes the end. Following Devlin et al.\cite{DBLP:conf/naacl/DevlinCLT19}, we use the hidden state of [CLS] as the representation of the sentence: $\textbf{S}=h_{[CLS]}$. 

The stylistic-based ranking score $Score_e(\tilde{S})$ for the distinct sentence $\tilde{S}$ is defined as below:
\begin{equation}\label{inter2}
\begin{aligned}
Score_e(\tilde{S}) = softmax(\textbf{S} \cdot W),
\end{aligned}
\end{equation}
where $W \in R^{H x 2}$ represents trainable weights and $H$ is the dimention of hidden state. 

\subsection{Diversity-Weighted Re-ranking}
\label{sec: diversity}

Above scores reflect the representativeness of
whether a review sentence is a tip or not in two different aspects. We combine the scores as below:
\begin{equation}\label{hybrid}
\begin{aligned}
Score(\tilde{S}) = Score_a(\tilde{S}) + \alpha \cdot Score_e(\tilde{S}),
\end{aligned}
\end{equation}
where $Score(\tilde{S})$ is the final ranking score
for a review sentence, and $\alpha$ is a parameter to adjust the contribution of the two aspects for the final ranking.

Although highly-ranked
sentences are representative, there are lots of redundant phrases in top sentences ordered by the above $Score$.
Considering that users do not want to see similar content frequently,
we propose a re-ranking method in this section. 
When re-ranking, both relevance and diversity factors should be considered.

\begin{figure}[ht]
\centering
\includegraphics[width = 0.48\textwidth]{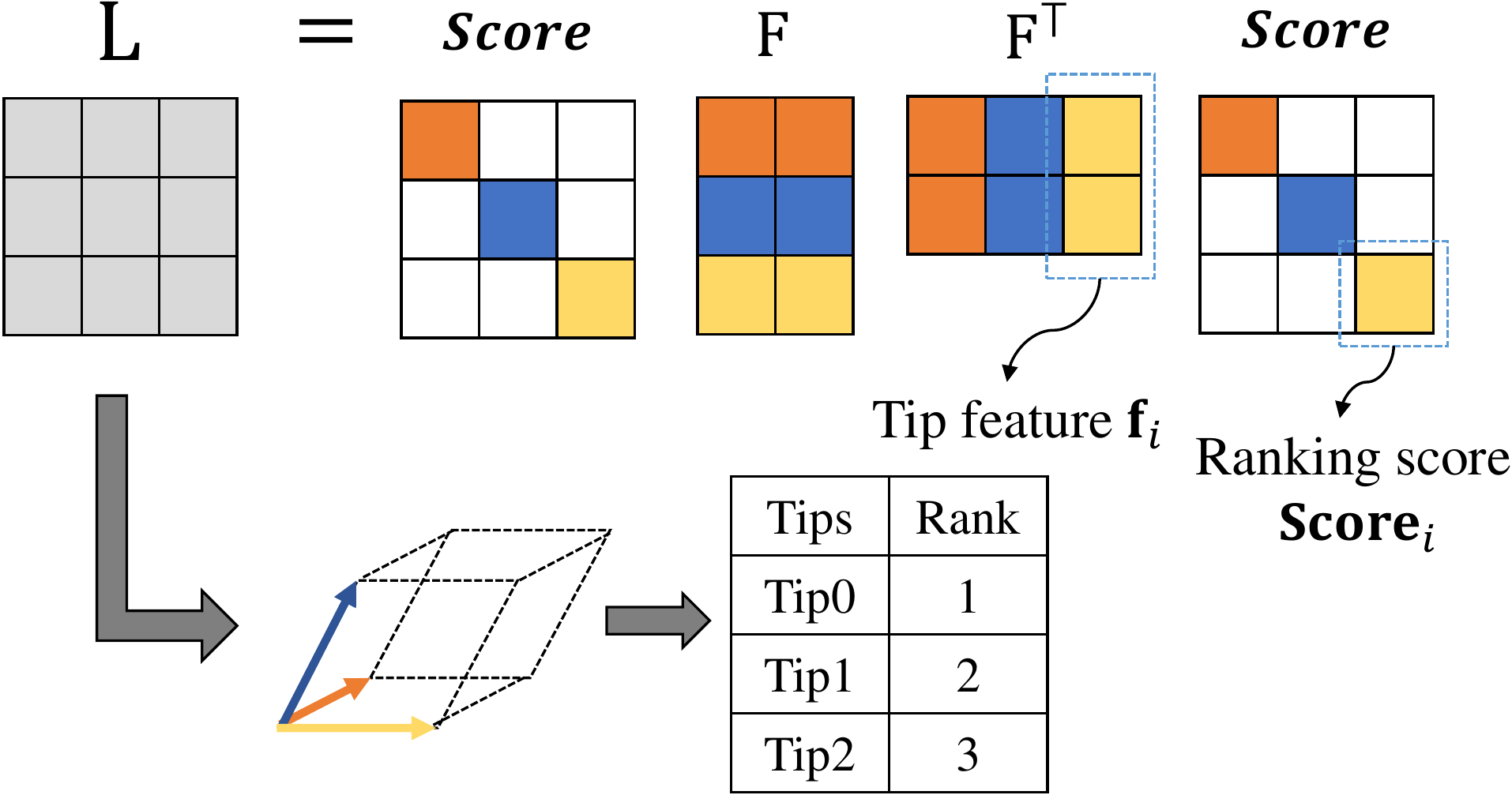}
\caption{Diversity-weighted re-ranking.}
\label{diversity}
\end{figure}

Figure~\ref{diversity} demonstrates how we adapt DPP~\cite{DBLP:journals/ftml/KuleszaT12} model in tip re-ranking. DPP selects a set of most representative sentences from candidates as tips. 
We first need to construct a kernel matrix $L=A^TA$, where the columns of $A$ are vectors representing the tips. 
To capture the topic of each candidate tip, we simply employ the typical topic modeling approach, PLSA~\cite{hofmann2013probabilistic}, to embed a candidate tip as $\textbf{f}$.
We can construct each column vector $A_i$ as the product of
the ranking score $Score(\tilde{S}_i)$ and the
feature vector $\textbf{f}_i$.
The entries of $L$ can be 
denoted
as
\begin{equation}
    \label{eq1} 
    \begin{aligned}
    L_{ij}&= \langle Score_i \cdot \textbf{f}_i, Score_j \cdot \textbf{f}_j \rangle \\
    &=Score_i \cdot \langle \textbf{f}_i,\textbf{f}_j \rangle \cdot Score_j  \\
    \end{aligned}
\end{equation}
where $i, j$ denote the indices of candidate tips. $ \langle \cdot,\cdot \rangle$ is the dot product of two vectors. 
The kernel matrix can be written as 
\begin{equation}
    \label{div2} 
    \begin{aligned}
    L = Diag(Score) \cdot F \cdot F^T \cdot Diag(Score),
    \end{aligned}
\end{equation}
\noindent where $F \cdot F^T $ denotes the similarity matrix and $Diag(Score)$ is a diagonal matrix whose diagonal element is $Score_i$.

Following Chen et al.~\cite{DBLP:conf/nips/ChenZZ18}, we use the fast greedy MAP inference algorithm for DPP.
$Y$ is the output candidate set, which is initialized as empty and added with the $j$-th sentence
\begin{equation}\label{eq4}
j=\mathop{\arg\max}_{i\in D \setminus Y} \log det(L_{Y \cup i} )-\log det(L_{Y})
\end{equation}
in each iteration.
We set the number of selected tips in $Y$ as 1, 3, 5, 10 for different experimental setup, respectively.


\section{Experiment Setup}
\label{sec:setup}
\subsection{Dataset}
The \dataset dataset is described in section~\ref{sec:Experimental}.
We split the dataset into training, validation and test sets. 
We first randomly select 9 songs and 50 sentences from each of them to form the test sets. This is to test the model's ability to recognize new tips under new reviews of the seen songs. 
Then we randomly split the remaining part into training and validation sets in the ratio of 8:2. 

Besides, to stimulate the practical use of tip extraction,
we evaluate our framework on previously-unseen songs. Specifically, we select 500 reviews from 9 unseen songs and split them into sentences based on the punctuation model described in Section~\ref{subsec:data}. The number of sliced sentences in each song exceeds 1,000, which is larger than that in our dataset \dataset, but more closer to real scenario.
Finally, the annotators manually evaluate the quality of the tips. Each sentence can be considered as a tip only if two annotators agree on it.
\subsection{Baselines}
\label{sec:baseline}
We compare our framework with several baseline classifiers as follows:

\textbf{SVM}~\cite{DBLP:conf/ecml/Joachims98} is a traditional machine learning model. Sentence embeddings are generated by Word2vec~\cite{mikolov2013efficient} which is trained by SogouNews corpus~\cite{li2018analogical}.
The size of embedding vector is 300. We rank tips according to the output probabilities.

\textbf{FastText}~\cite{DBLP:conf/eacl/GraveMJB17} is a shallow neural network and uses n-gram features additionally to capture partial information. It has fast convergence speed for training and high speed for inference. We randomly initialize tokens to 300-dimensional embeddings. We rank tips by the classification score. 

\textbf{BiLSTM-Attention}~\cite{DBLP:conf/naacl/YangYDHSH16} exploits the bidirectional long short term memory cell mechanism and
uses 
attention mechanism to alleviate long-range dependencies. We initialize token embeddings to 300-dimensional vectors with Word2Vec trained by SogouNews corpus. We rank tips by the classification score. 

\textbf{BERT}~\cite{DBLP:conf/naacl/DevlinCLT19} is short for Bidirectional Encoder Representations from Transformer. We use the pre-trained `bert-base-chinese' model~\cite{DBLP:conf/emnlp/WolfDSCDMCRLFDS20} to fine tune a binary classifier for our task. The first token ([CLS]) representation is fed into an output layer for classification. We rank tips by the classification score. 

\subsection{Implement Details}
We use $BERT_{base}$ for Chinese from Transformers~\cite{DBLP:conf/emnlp/WolfDSCDMCRLFDS20}. 
we apply AdamW~\cite{DBLP:journals/corr/abs-1711-05101}($\beta_1$=0.9, $\beta_2$=0.999) with a weight decay rate of 0.01 and set the learning rate to 2e-5.
The damping coefficient $d$ (in Eq.~\ref{transfer eq}) is defined as 0.85.
We set the base number $m$ as 10 to smooth 
the approval number when constructing sentence graph in Content-based Ranking.
The parameter $\alpha$ (in Eq.~\ref{hybrid}) for adjusting
the ranking score is defined as 0.8 and the topic number $k$ in PLSA is set as 8. Detailed analysis about the parameter settings will be introduced in Section~\ref{sec: parameter analysis}.

\section{Experiment Evaluation}
\label{sec:exper}
In the following section, we present evaluation methods for our framework. 
We firstly analyze performance of the whole framework
. Then we report the detailed analysis of each module.

\subsection{Results and Discussion}
\begin{table}[t]
\caption{Precision of Top-k in the test set.}
\label{seen}
\centering
\scalebox{0.9}{
    \begin{tabular}{c|cccc}
    \toprule[1.5pt]
    Model & p@1 & p@3 & p@5 &p@10\\
    \midrule[1pt]
    SVM &44.44&70.37&73.33&75.56\\
    FastText &66.67&74.07&66.67&72.22\\
    BiLSTM-Attn &33.33&62.96&64.44&66.67\\
    BERT &77.78&88.89&84.44&82.22 \\
    
    \hline
    
    \multicolumn{1}{c|}{Ours} & \textbf{77.78} & 85.19 & \textbf{86.67} & \textbf{85.56} \\
    \multicolumn{1}{c|}{w/o Content-based ranking}& 77.78 & 85.19 & 86.67 & 82.22 \\
    \multicolumn{1}{c|}{w/o Stylistic-based ranking} & 33.33 & 51.85 & 57.78 & 52.22 \\
    \multicolumn{1}{c|}{ \makecell[c]{w/o Diversity-weighted \\ re-ranking}} & 77.78 & \textbf{88.89} & 86.67 & 84.44 \\
    
    \bottomrule[1.5pt]
    \end{tabular}
}
\end{table}

\begin{table}[t]
\caption{Precision of Top-k in unseen songs.}
\label{unseen}
\centering
\scalebox{0.9}{
    \begin{tabular}{c|cccc}
    \toprule[1.5pt]
    
     & p@1 & p@3 & p@5 &  p@10\\
    
    \midrule[1pt]
    
    SVM & 66.67 & 59.26 & 60.00 & 57.78\\ 
    FastText & 66.67 & 66.67&57.78 & 56.67\\
    BiLSTM-Attn & 44.44 & 55.56 & 53.33 & 53.33\\
    BERT & 77.78 & 81.48 & 82.22 & 78.89\\
    
    \hline
    
    \multicolumn{1}{c|}{Ours} & \textbf{88.89} & \textbf{88.89} & \textbf{86.67} & \textbf{78.89} \\
    \multicolumn{1}{c|}{w/o Content-based ranking} & 88.89 & 85.19 & 82.22 & 76.67 \\
    \multicolumn{1}{c|}{w/o Stylistic-based ranking} & 44.44 &	40.74 &	40.00 & 38.89 \\
    \multicolumn{1}{c|}{\makecell[c]{w/o Diversity-weighted \\ re-ranking}} & 88.89 &	85.19 &	82.22 &	78.89 \\
    
    \bottomrule[1.5pt]
    \end{tabular}
}
\end{table}

We compare \tool with the four baseline models introduced in Section~\ref{sec:baseline} on \dataset. The comparison results are shown 
 in Table~\ref{seen} and Table~\ref{unseen}. 
Based on the comparison results, we obtain 
 the following observations:

\textbf{Observation 1: \tool 
outperforms all the baseline models.} As 
shown in Table~\ref{seen} and Table~\ref{unseen}, \tool presents better performance than the corresponding baselines on both seen and unseen songs. 
For example. \tool improves BERT by 2.23\%, 4.45\% respectively for seen songs and unseen songs in terms of the precision of top-5. The results demonstrate that with content- and stylistic-pattern joint, \tool can more accurately extract tips from song reviews. 

\textbf{Observation 2: Stylistic-based ranking plays
a key role in tip identification.} 
Comparing \tool w/o content-based ranking with \tool, we can find that \tool performs worse without the stylistic-based ranking component, even worse than the
baseline models. For example, the performance of \tool drops by 28.89\% without the stylistic-based ranking module for seen songs in terms of the precision of top-5.
The phenomenon indicates that stylistic pattern is more important in tips identification. 

\textbf{Observation 3: Models' performance shows difference for seen songs and unseen songs.}
During evaluating on seen songs, we try to recognize new tips for the songs appearing
in the training set. 
For the evaluation on unseen songs, we try to identify tips for songs which are not in training set. 
Comparing the prediction results in the two settings, as shown in Table~\ref{seen} and Table~\ref{unseen}, the performance of baseline models generally drops with respect to different metrics for the unseen songs. For example, the performance of BERT for unseen songs is 2.23\% lower than that for seen songs; FastText drops 8.89\% in unseen songs compared to seen songs. 
This results are reasonable since the prior knowledge is limited when evaluating on the unseen songs, and is more challenging. However, the proposed \tool is relatively more stable for both settings. The results further indicate the effectiveness of \tool in tip generation.

\subsection{Parameter Analysis}
\label{sec: parameter analysis}
In the section, we analyze the impact of two important hyper-parameters on the performance of \tool, including the base number $m$ for smoothing approval numbers in content-based ranking module and the topic number $t$ in diversity-weighted re-ranking module. 
\subsubsection{Approval Number Smoothing}
\begin{figure}[ht]
\centering
\includegraphics[width = 0.45\textwidth]{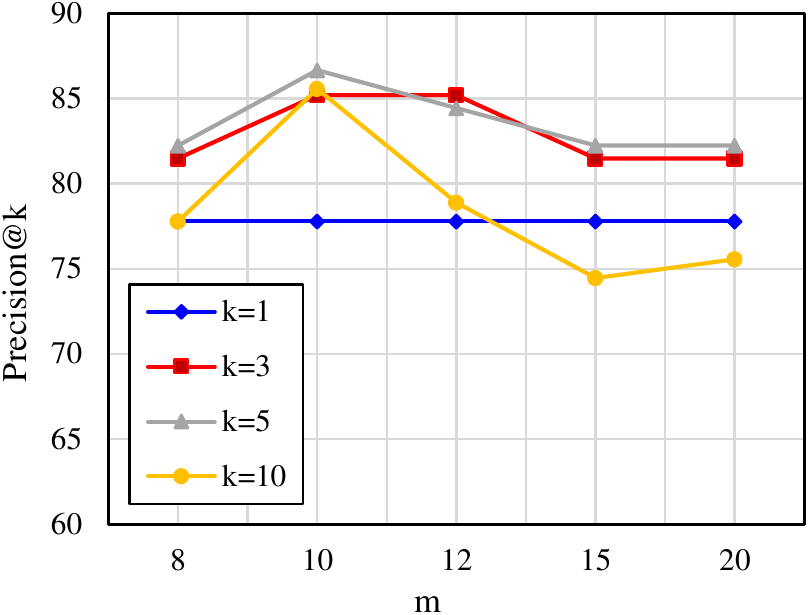}
\caption{Precision@top-k of \tool with approval numbers smoothed by different base number $m$ in log function.
}
\label{base_num}
\end{figure}

In content-based ranking module, we take some strategies to smooth the approval numbers of reviews. Approval numbers conform 
to the long-tailed distribution, that is to say, the approval numbers from top reviews occupy the majority of the approval numbers from all reviews. Since the reviews with few approval numbers can also be representative as tips, e.g., some fresh reviews, the extreme imbalance distribution should be alleviated.
We choose logarithmic function 
and perform detailed experiment to select the base number of the log function, with the results shown in Figure~\ref{base_num}. As can be seen, \tool can achieve the best performance with respect to most of the metrics when $m$ is set as 10. Thus, we choose $m=10$ for evaluation during experiments.
\begin{figure}[ht]
\centering
\includegraphics[width = 0.45\textwidth]{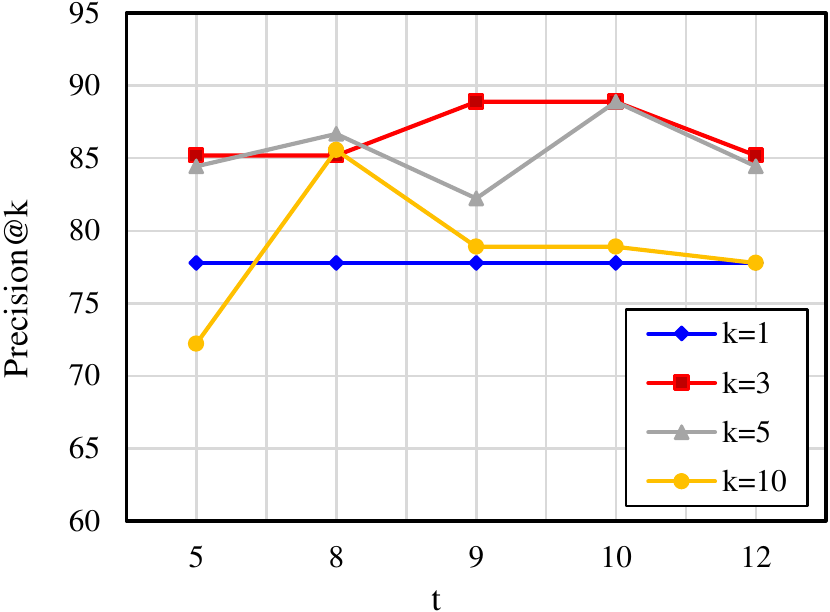}
\caption{Precision@top-k of \tool along with
different numbers of topics $t$ in PLSA.}
\label{topic_num}
\end{figure}

\subsubsection{Topic Number}

To represent each tip as a feature vector $\textbf{f}$, we deploy topic model (PLSA) in the diversity-weighted re-ranking module, as illustrated in Section~\ref{sec: diversity}. We analyze the impact of different topic number $t$ on the performance of \tool. As can be seen in Figure~\ref{topic_num}, the performance of \tool generally fluctuates along with the increase of $t$. Besides, smaller or larger $t$ cannot bring much improvement on the prediction results. Since \tool achieves relatively better performance regarding all the metrics when $t$ equals 8, we set $t=8$ during experimental evaluation.

\begin{table*}[t]
\scriptsize
\centering
\caption{The metric \textit{Distinct} of top-k tips, k=3,5,10.}
\label{distinct}
\scalebox{1.2}{
\begin{tabular}{ccccccc}
\toprule
\multirow{2}{*}{Model} & \multicolumn{2}{c}{Top-3} & \multicolumn{2}{c}{Top-5} & \multicolumn{2}{c}{Top-10} \\
\cmidrule(r){2-3} \cmidrule(r){4-5} \cmidrule(r){6-7}
& Distinct-1      &  Distinct-2 
& Distinct-1     &  Distinct-2   
& Distinct-1      &  Distinct-2  \\
\midrule
\tool
& \textbf{83.2} & \textbf{96.7}  & \textbf{78.9} & \textbf{95.3} & \textbf{73.7} & \textbf{93.2} \\
w/o diversity-weighted re-ranking & 78.0 & 93.4  & 75.3 & 92.8 & 72.0 & 91.8  \\
\bottomrule

\end{tabular}
}
\end{table*}

\begin{table*}[ht]
\caption{Results of top-5 tips with \tool and \tool w/o diversity-weighted re-ranking}
\label{case_study}
\centering
\scalebox{0.9}{
\begin{tabular}{ccc}
\toprule[1.5pt]
Rank & \tool & \tool w/o diversity-weighted
\\
\midrule[1pt]
1  
& \makecell[c]{\begin{CJK*}{UTF8}{gbsn}日本真是一个可爱可喜的国家\end{CJK*} \\ (Japan is such a lovable and delightful country)} 
& \makecell[c]{\begin{CJK*}{UTF8}{gbsn}来自\colorbox{yellow}{深海的少女}\end{CJK*} \\ (The girl from the deep sea)} \\
\midrule[0.5pt]
2 
& \makecell[c]{\begin{CJK*}{UTF8}{gbsn}谁又说过客就不是命中注定\end{CJK*} \\ (Who says that passing visitors are not destined)}
& \makecell[c]{\begin{CJK*}{UTF8}{gbsn}岛风沉入海里面就变成了\colorbox{yellow}{深海少女}\end{CJK*} \\ (Shimakaze sinks into the sea and becomes a deep sea girl)} \\

\midrule[0.5pt]

3 
& \makecell[c]{\begin{CJK*}{UTF8}{gbsn}岛风沉入海里面就变成了\colorbox{yellow}{深海少女}\end{CJK*} \\ (Shimakaze sinks into the sea and becomes a deep sea girl)}
&\makecell[c]{\begin{CJK*}{UTF8}{gbsn}\colorbox{yellow}{深海少女}，是谁惹红了你的双眼\end{CJK*} \\ (Deep sea girl, who has made your eyes red)} \\

\midrule[0.5pt]

4 
& \makecell[c]{\begin{CJK*}{UTF8}{gbsn}很带感的歌，适合单曲循环\end{CJK*} \\ (very emotional song, suitable for single song cycle)}
& \makecell[c]{\begin{CJK*}{UTF8}{gbsn}我在\colorbox{yellow}{深海}等你\end{CJK*} \\ (I am waiting for you in the deep sea)} \\

\midrule[0.5pt]

5 
& \makecell[c]{\begin{CJK*}{UTF8}{gbsn}\colorbox{yellow}{深海少女}，是谁留下了你的纯洁\end{CJK*} \\ (Deep sea girl, who left your purity)} 
& \makecell[c]{\begin{CJK*}{UTF8}{gbsn}来自\colorbox{yellow}{深海的少女}\end{CJK*} \\ (The girl from the deep sea)}\\

\bottomrule[1.5pt]
\end{tabular}
}
\end{table*}

\subsection{Diversity Analysis}
We adopt the evaluation metric named \textit{Distinct}~\cite{DBLP:conf/naacl/LiGBGD16} from text generation task to measure the diversity of top-k tips.
Specifically, $Distinct\text{-}1$ and $Distinct\text{-}2$ are the proportion of distinct unigrams and bigrams in the total number of the corresponding generated unigram or bigrams, respectively. 


We compare the results of the metric \textit{Distinct} for \tool and \tool w/o diversity-weighted re-ranking, as illustrated in Table~\ref{distinct}.
It can be seen that \tool consistently outperforms \tool without the diversity-weighted re-ranking module for different numbers of top tips. For example, for the top-5 tips, \tool increases by 3.6\% and 3.5\% in terms of $Distinct$-1 and $Distinct$-2, respectively. The results indicate that the diversity-weighted re-ranking module is helpful for generating tips with diverse topics. We also illustrate the results of the two approaches for the song ``Sea girl''(\begin{CJK*}{UTF8}{gbsn}``深海少女''\end{CJK*}) in Table~\ref{case_study}. As can be seen, the top-5 tips ranked by \tool without the diversity-weighted re-ranking module deliver similar topic ``Sea girl''. However, the tips generated by \tool are more diverse in topic. For example, the first tip ranked by \tool talks about the country related to the song, and the fourth tip discusses about the song rhythm.